# Dirac neutrinos in the seesaw mechanism. Dirac lepton number violation

Igor T. Dyatlov *
Scientific Research Center "Kurchatov Institute"
Petersburg Institute of Nuclear Physics, Gatchina, Russia

The seesaw mechanism explains the exclusive smallness of neutrino masses by the presence of very heavy Majorana masses and leads to the appearance of Majorana particles and to the direct lepton number violation. The author proposes a seesaw scenario that produces only Dirac neutrinos with the same violation. This scenario appears possible for heavy neutrinos with non-perturbative Higgs boson *H* couplings. Such a scenario could be accomplished in the model describing the structure of weak mixing matrices for quarks and leptons through the existence of very heavy mirror analogs of Standard Model fermions. The non-perturbativity of the problem hinders the analytical solution, but the conditions derived indicate that the mechanism under consideration preferentially generates only Dirac neutrinos. This phenomenon may have relevance for leptogenesis processes if all existing neutrinos turn out to be of the Dirac type.

## 1. Introduction

Majorana masses violating the lepton number are an essential element of the seesaw mechanism—a possible explanation for exceptionally small neutrino masses. The seesaw model is well known, widely used, and extensively described (see reviews [1]):it results in two neutrinos, one heavy and one light, that must be Majorana particles.

The Lagrangian of the seesaw mechanism for neutrino masses (one generation) is written out as:

$$\mathcal{L}^{(\nu)} = \mu\left(\bar{\Psi}_R\Psi_L + \bar{\Psi}_L\Psi_R\right) + \frac{M_R}{2}\left(\Psi_R^T C \Psi_R + \bar{\Psi}_R C \bar{\Psi}_R^T\right), \tag{1}$$

where $\Psi_{R,L}$ are the neutrino's chiral components (R, L), $\Psi_R$ is a weak isoscalar, and $\Psi_L$ is an isospinor component. The Lagrangian includes one Dirac ($\mu$) and one Majorana ($M_R$) mass. When choosing (1), the weak $SU_L(2)$ symmetry is a central factor to consider. The Dirac mass $\mu$ appears in the Standard Model (SM) as a result of the spontaneous violation of $SU_L(2)$ by the vacuum average of the Higgs isodoublet $\Phi_H$ [2]. The Majorana mass $M_R$ can be present in the Lagrangian itself, since $\Psi_R$ is an isoscalar. It can also be produced by a process similar to the production of $\mu$ with the vacuum average of the new isoscalar meson.

The Majorana terms with the $\Psi_L$ isodoublet component are not generally considered. To introduce them without violating $SU_L(2)$-invariance requires a completely different, more complicated procedure [1]—the existence of isovector scalars with their own vacuum averages or non-renormalizable terms $\sim \Phi_H^2$. However, even the new arbitrary constants ($M_L$) do not change the Majorana character of the resultant neutrinos.

*E-mail: dyatlov@thd.pnpi.spb.ru

It seems obvious that the seesaw mechanism does not allow a combination of an acceptable, suitable weak $SU_L(2)$ symmetry violation with Dirac states of neutrinos, i.e., the simultaneous presence of the symmetrical $\Psi_R$ and $\Psi_L$ components.

In the phenomenological model proposed by the author [3,4], the observed qualitative structures of weak mixing matrices (WMM) for quarks and leptons are explained through the existence of very heavy mirror analogs of SM fermions (the mirror symmetry violation model). For neutrinos, the model reproduces the observed qualities; namely, the exceptionally small masses and the considerable difference of the lepton WMM from the one for quarks [5], including such details as the smallness of the mixing angle sine $\Theta_{13}$. Reproduction is possible here if both heavy (mirror) and light (SM) neutrinos are of the Dirac type. This could be achieved in a seesaw mechanism analog with specifically fitted Majorana constants ($M_R$, $M_L$). Even at $\mu \ll M$, this seesaw model generates a single (Dirac) state with a specific mass for each of the light and heavy (mirror) particle systems, rather than two (Majorana) states with different masses.

This paper investigates how the desired system of Dirac states can be generated in the seesaw scheme through a dynamical mechanism triggered by the necessary presence in the mirror model of the non-perturbative coupling between heavy mirror $\Psi$ (not yet the model's physical neutrinos) and the Higgs boson *H*. Only a non-perturbative mechanism can generate the mass $M_L$. A direct weak $SU(2)$ symmetry violation, by introducing $M_L$ immediately into the Lagrangian (1), does not happen here. Since *H* appears in the broken $SU(2)$ system, the model under consideration is a consequence of the usual Higgs mechanism of violation [2] and is impossible in the system without symmetry breaking. In the mirror scheme, the weak $SU(2)$ symmetry becomes chiral: $SU(2) \rightarrow SU_L(2)$ only upon mirror symmetry breaking. This breaking means that the weak $SU(2)$ symmetry is violated too. In such a system of heavy Dirac neutrinos, lepton number violation would occur in interactions with the *H* boson and weak *W* vecton.

The non-perturbativity of the system hinders the analytical consideration of the process. We have to limit our investigation to the properties of the basic equation and conditions for the existence of its solutions. These conditions correspond precisely to Dirac-type neutrinos.

The coupling of physical heavy mirror neutrinos with the Higgs boson *H* is obscured here by non-perturbativity. This coupling may disobey the SM rule that it be simply proportional to particle masses. This situation is already known to exist even in the usual (perturbative) seesaw system (see Eq.(11), Eq.(12)). At the same time, the constants of the *H* coupling with light neutrinos, which in our model represent SM neutrinos, are proportional to their masses. Therefore, the SM perturbative properties do not change for the observed particles with their masses. In particular, owing to cancellation of W contributions by contributions involving H, cross-sections of processes that produce longitudinally polarized vector W bosons do not increase rapidly [2].

In Section 2, the usual seesaw mechanism is compared with the proposed model involving Dirac neutrinos. A discussion of the properties and differences of the two systems is provided. In Section 3, we derive an approximate equation that can produce Dirac particles, and find conditions for the existence of its solutions. Section 4 discusses the relation of the heavy neutrino mechanisms under consideration with SM light neutrino physics.

## 2. Majorana mass terms for Majorana and Dirac neutrinos

To clearly understand the discussion that follows, let us look again at the procedure of Majorana state generation in the Lagrangian (1). This procedure, in this or that form, is present in all representations of the seesaw mechanism [1]. Let us introduce the following operators:

$$\chi_R = \frac{\Psi_R + C\bar{\Psi}_R^T}{\sqrt{2}}, \quad \chi_L = \frac{\Psi_L + C\bar{\Psi}_L^T}{\sqrt{2}}, \tag{2}$$

where C is the charge conjugation matrix, which is assumed to be real $C = -C^T, C^2 = -1$. In the terms of (2), the Lagrangian (1) is written out as:

$$\mathcal{L}^{(\nu)} = \mu\left(\bar{\chi}_R \chi_L + \bar{\chi}_L \chi_R\right) + M_R(\bar{\chi}_R \chi_R). \tag{3}$$

Diagonalization (3) is the diagonalization of the matrix

$$\begin{array}{cc} \bar{\chi}_R & \bar{\chi}_L \end{array}$$

$$\begin{vmatrix} M_R & \mu \\ \mu & 0 \end{vmatrix} \begin{array}{c} \chi_R \\ \chi_L \end{array}. \tag{4}$$

It results in the eigenvalues

$$\lambda_\pm = \frac{M_R}{2} \pm \sqrt{\frac{M_R^2}{4} + \mu^2} \tag{5}$$

and eigenfunctions of Majorana fermions

$$\chi_+ = \frac{1}{\sqrt{N}}\left[\chi_R - \frac{\mu}{\lambda_+}\chi_L\right],$$
$$\chi_- = \frac{1}{\sqrt{N}}\left[\chi_L + \frac{\mu}{\lambda_+}\chi_R\right], \quad N = \frac{\lambda_+^2}{\lambda_+^2 + \mu^2}. \tag{6}$$

At $M_R \gg \mu$, we obtain the well known [1], very large and very small masses of the Majorana states (6)

$$\lambda_+ \simeq M_R, \quad \lambda_- \simeq -\frac{\mu^2}{M_R}. \tag{7}$$

Of particular interest are the interactions of the neutrino (6) with the Higgs scalar doublet $\Phi_H$ and vector boson *W*. The coupling between these interactions, ensuing from weak $SU_L(2)$ symmetry spontaneous violation:

$$\langle\Phi_H\rangle = \frac{\eta}{\sqrt{2}}, \quad \eta \simeq 246\,\text{ГэВ}\,[5], \tag{8}$$

results, in SM, in the cancellation of contributions from diagrams with *H* and *W* participation, limiting the growth of cross-sections with energy. In addition, in the invariant gauge, contributions from the pole $q^2 = 0$ in the propagator *W*

$$\frac{g_{\mu\nu} - q_\mu q_\nu/q^2}{q^2 - M_W^2} = \frac{g_{\mu\nu} - q_\mu q_\nu/M_W^2}{q^2 - M_W^2} + \frac{q_\mu q_\nu}{q^2}\frac{1}{M_W^2}, \tag{9}$$

are cancelled by the Goldstone boson contribution $\varphi = \eta\theta/2$ of the violated symmetry:

$$\Phi_H = \frac{\eta + H}{\sqrt{2}}\,e^{i(\vec{\theta}\vec{\tau})/2}, \tag{10}$$

where $\vec{\tau}$ are isospin matrices. For interactions with neutrinos only, $\vec{\tau} \to \tau_3 \to 1$. Contributions of diagrams of the same order in the perturbation theory cancel out. This perturbative cancellation becomes possible if constants of the *H* boson coupling with fermions are small.

In SM without seesaw, these properties are achieved as a result of $\mu$ essentially being the fermion mass. In the seesaw system, at $\lambda_\pm$ masses (5), interactions of $\Psi_L$ and $\Psi_R$ with $\Phi$ are not proportional to the masses

$$\frac{\sqrt{2}\mu}{\eta}\left(\bar{\Psi}_L\Psi_R\frac{\eta+H}{\sqrt{2}}e^{i(\theta\tau/2)} + \bar{\Psi}_R\Psi_L\frac{\eta+H}{\sqrt{2}}e^{-i(\theta\tau/2)}\right) \Rightarrow$$
$$\Rightarrow \mu(\bar{\Psi}\Psi) + \frac{\mu}{\eta}(\bar{\Psi}\Psi)H + i\frac{\mu}{\eta}(\bar{\Psi}\gamma_5\Psi)\varphi. \tag{11}$$

The interaction with the Higgs boson *H* has the form:

$$\frac{\mu}{\eta}\bar{\Psi}\Psi H = \frac{\mu}{\eta N}\left[-\frac{2\mu}{\lambda_+}\bar{\chi}_-\chi_- + \frac{2\mu}{\lambda_+}\bar{\chi}_+\chi_+ + \frac{M_R}{\lambda_+}\left(\bar{\chi}_+\chi_- + \bar{\chi}_-\chi_+\right)\right]H. \tag{12}$$

Here, only the interaction with $\chi_-$ is proportional to its mass

$$-\frac{\mu^2}{\lambda_+} = \frac{\lambda_+\lambda_-}{\lambda_+} = \lambda_-. \tag{13}$$

How do cancellations of contributions occur in the seesaw system? The coupling with the Goldstone "particle" $\varphi$ in the terms of (2) and (6) assumes the form similar to (12):

$$i\frac{\mu}{\eta}(\bar{\Psi}\gamma_5\Psi)\varphi = i\frac{\mu}{\eta N}\left[\frac{M_R}{\lambda_+}\left(\bar{\chi}_+\gamma_5\chi_- + \bar{\chi}_-\gamma_5\chi_+\right) + \frac{2\mu}{\lambda_+}\bar{\chi}_+\gamma_5\chi_+ - \frac{2\mu}{\lambda_+}\bar{\chi}_-\gamma_5\chi_-\right]\varphi. \tag{14}$$

The interaction of neutrinos (6) with a pole of the propagator $W$ in (9) appears to be exactly the same ($M_W = \frac{1}{2}g_2\eta$, [2]):

$$g_2\frac{q^\mu}{M_W}\left(\bar{\Psi}_L\gamma_\mu\frac{\tau\varphi}{2}\Psi_L\right) \Rightarrow \frac{\mu}{\eta N}\left[\frac{M_R}{\lambda_+}\left(\bar{\chi}_+\gamma_5\chi_- + \bar{\chi}_-\gamma_5\chi_+\right) + \frac{2\mu}{\lambda_+}\bar{\chi}_+\gamma_5\chi_+ - \frac{2\mu}{\lambda_+}\bar{\chi}_-\gamma_5\chi_-\right]\varphi \tag{15}$$

$$q = p_1 - p_2, \quad \hat{p}\chi_\pm = \lambda_\pm\chi_\pm.$$

Owing to the connection of contributions with the boson $H$ and contributions with weak interaction (similar to SM), cross-sections with participation of longitudinal $W$ do not increase fast.

The analysis above is applicable to the perturbative theory coupling $\mu/\eta \ll 1$. Large constants alter the mechanism significantly.

Let us now contemplate a system which, while including a Majorana mass term, results in Dirac particles. Such a system was constructed in [2,3] for a mirror symmetry model reproducing WMM properties for quarks and leptons. The mass term of the Lagrangian for that model is

$$\mathcal{L}^{(\nu)} = \mu\left(\bar{\Psi}_R\Psi_L + \bar{\Psi}_L\Psi_R\right) + \frac{M}{2}\left(\Psi_R^T C\Psi_R - \Psi_L^T C\Psi_L\right) + c.c. \tag{16}$$

Only at this sign between the equal $L$ and $R$ terms, Eq.(16) represents a Dirac neutrino; otherwise it is a Majorana one.

The Lagrangian (16) differs from (1) in that it has a term with the isospinor $\Psi_L$ and violates the $SU(2)$ invariance. In addition, as it has already been mentioned in Section 1, this term cannot simply be obtained from the $SU(2)$ invariant Lagrangian; it is necessary to introduce violation using vacuum averages of the isovector scalar or employ non-renormalizable interactions of fermions with the square of the Higgs scalar, $\sim \Phi_H^2$ [1]. However, in those conditions the equality of the factors $M_R = -M_L$ does not appear to have any grounds.

A solution for this problem will be proposed later in this paper; in the meantime, let us describe the properties of the neutrino system defined by Eq.(16). The Lagrangian (16) can be rewritten in terms of Dirac operators:

$$\Psi = \Psi_R + \Psi_L. \tag{17}$$

Then we have:

$$\mathcal{L}^{(\nu)} = \mu \bar{\Psi}\Psi + \frac{M}{2}\left(\bar{\Psi}^C \gamma_5 \Psi - \bar{\Psi}\gamma_5 \Psi^C\right), \tag{18}$$
$$\Psi^C = C\bar{\Psi}^T.$$

Diagonalization of this expression does result in Dirac operators [4]:

$$\Psi_\lambda = \frac{1}{\sqrt{2N}}\left[\left(1 - \frac{\mu}{M+\lambda}\right)\gamma_5 \Psi + \left(1 + \frac{\mu}{M+\lambda}\right)\Psi^C\right],$$
$$\Psi_\lambda^C = \frac{1}{\sqrt{2N}}\left[\left(1 + \frac{\mu}{M+\lambda}\right)\Psi - \left(1 - \frac{\mu}{M+\lambda}\right)\gamma_5 \Psi^C\right], \tag{19}$$
$$\lambda = (M^2 + \mu^2)^{1/2}, \quad N = \frac{2\lambda}{M+\lambda}, \quad \Psi_\lambda \neq \Psi_\lambda^C.$$

The Lagrangian (18) acquires the form:

$$\mathcal{L}^{(\nu)} = \lambda \bar{\Psi}_\lambda \Psi_\lambda, \tag{20}$$

i.e., becomes a Dirac mass term. This is also supported by the transition to $\Psi_\lambda$ in the kinetic term of the Lagrangian (see [4b]).

The usual purpose of seesaw problems to explain the appearance of very small particle masses apparently cannot be achieved in this scenario. But in our model [1,2] the existence of very heavy Dirac mirror neutrinos is precisely what leads to exceptionally small masses of Dirac neutrinos in SM and suitable qualitative properties of the WMM.

Inversion of Eq.(19)

$$\Psi = \frac{1}{\sqrt{2N}}\left[\left(1 - \frac{\mu}{M+\lambda}\right)\gamma_5 \Psi_\lambda + \left(1 + \frac{\mu}{M+\lambda}\right)\Psi_\lambda^C\right],$$
$$\Psi^C = \frac{1}{\sqrt{2N}}\left[\left(1 + \frac{\mu}{M+\lambda}\right)\Psi_\lambda - \left(1 - \frac{\mu}{M+\lambda}\right)\gamma_5 \Psi_\lambda^C\right] \tag{21}$$

allows us to determine the interaction with the Higgs boson. It involves terms that do not conserve the lepton number:

$$\frac{\mu}{\eta}(\bar{\Psi}\Psi)H = \left[\frac{\mu^2}{\lambda\eta}(\bar{\Psi}_\lambda\Psi_\lambda) + \frac{\mu M}{2\lambda\eta}\left(\bar{\Psi}_\lambda^C\gamma_5\Psi_\lambda - \bar{\Psi}_\lambda\gamma_5\Psi_\lambda^C\right)\right]H. \tag{22}$$

The coupling $\Psi_\lambda$ with the "Goldstone" particle of the violated symmetry equals (neutrino interactions only: $\tau \to 1$)

$$i\frac{\mu}{\eta}\bar{\Psi}\gamma_5\Psi\varphi = i\left[\frac{\mu^2}{\lambda\eta}(\bar{\Psi}_\lambda\gamma_5\Psi_\lambda) + \frac{\mu M}{2\lambda\eta}\left(\bar{\Psi}_\lambda^C\Psi_\lambda - \bar{\Psi}_\lambda\Psi_\lambda^C\right)\right]\varphi,$$

$$\langle T(\varphi,\varphi)\rangle \sim \frac{1}{-q^2}. \tag{23}$$

At the same time, the interaction $\Psi_\lambda$ with a pole term in the *W* boson propagator (9) is

$$\frac{g_2}{2M_W}\bar{\Psi}_L\hat{q}\Psi_L\varphi_W = \frac{g_2}{2M_W}\left[\mu(\bar{\Psi}_\lambda\gamma_5\Psi_\lambda) + \frac{M}{2}\left(\bar{\Psi}_\lambda^C\Psi_\lambda - \bar{\Psi}_\lambda\Psi_\lambda^C\right)\right]\varphi_W,$$

$$\hat{p}\Psi_\lambda = \lambda\Psi_\lambda, \quad \langle T(\varphi_W,\varphi_W)\rangle = \frac{1}{-q^2}. \tag{24}$$

Assuming the spinor states $u$ and $v$, $\hat{p}\Psi_\lambda = \lambda\Psi_\lambda$, $\hat{p}\Psi_\lambda^C = -\lambda\Psi_\lambda^C$. In SM, the mass $M_W = g_2\eta/2$ and contributions from Eq.(23) and Eq.(24) are impossible to cancel out. This, however, is not unexpected due to an arbitrary incorporation of the $SU(2)$ symmetry violating term into the Lagrangian (16). Termwise, perturbative cancellation of contributions does not occur. We assert that the Dirac system (16) with lepton number violation could arise from the symmetry violating term obtained by a non-perturbative method, as a result of the strong interaction ($\mu/\eta \gg 1$) of the Higgs boson *H* with the neutrino. This situation exists for heavy mirror states in the model [3,4] (in the seesaw scenario $M/\eta \gg \mu/\eta \gg 1$).

Can the standard value of *W* boson mass $M_W = g_2\eta/2$ be changed in this scenario? Strong interactions related with participation of virtual heavy particles are confined here to a very small region: $1/\lambda \ll 1/M_W$. In [6] it is stated that the influence of these processes could be restricted by the volume ratio (an interaction region to process one), i.e., be very low. Corrections from virtual light fermions to the *W* mass will have the usual, standard value.

## 3. Non-perturbative transition of the Majorana system to Dirac

In the Lagrangian (1), the Dirac mass parameter $\mu$ appears in SM as a result of weak $SU(2)$ symmetry violation by the vacuum average of the Higgs doublet $\langle\Phi\rangle = \eta/\sqrt{2}$ (the mass $M_R$ does not violate the symmetry). The Higgs boson *H* (10) appears simultaneously. The *H* boson changes the chirality of neutrinos. This inevitably produces the Majorana mass $M_L$ in addition to the mass $M_R$ which is inherently present in the Lagrangian (as in (1)). Fig. 1 illustrates this phenomenon. At small Yukawa coupling constants, induced $M_L \neq M_R$ and fermions remain Majorana. At a large $\mu \gg \eta$, the interaction $\Psi$ with neutrinos becomes non-perturbative. The influence of this strong coupling on the Majorana mass is discussed below.

For this purpose, let us assume a simplified model involving the strong interaction $\Psi$ with $H$ ($\mu/\eta \gg 1$) and containing no fermion loops. Ultimately, it is anticipated that the fermions will be very heavy, which may weaken the influence of their loops on the processes under investigation (in the seesaw scenario, $M \gg \mu \gg \eta$).

Let us look at the equations for the coefficients

$$\text{Tr}\left(\bar{\Psi}^C_{R,L}(p)\,\Psi_{R,L}(p)\right) f_{R,L}(p), \tag{25}$$

that include all diagrams of the Fig. 2 type (Tr over spinor indices, $f_{RL}$ is defined as $\frac{1}{4}$Tr). The symbol $T$ in Fig. 2 denotes diagrams with an arbitrary number of $H$ lines that cannot be reduced down to splitting into parts joined by two fermion lines. The functions $f_R$ and $f_L$ are related to each other, since chiralities on fermion lines are changed both by the interaction with $H$ and by propagator mass terms. Of primary interest here is the possibility of formation of a Dirac neutrino, while the Lagrangian contains only $M_R$.

Fermion propagators $\langle T(\Psi, \overline{\Psi}) \rangle$ are calculated in the "mean field" approximation, i.e., in the form of expressions that we seek to obtain but with arbitrary, momentum-independent coefficients, defined by solution matching:

$$G(p) = \frac{\alpha + \beta \hat{p}}{\lambda^2 - p^2}, \quad \hat{p} = p_\mu \gamma^\mu. \tag{26}$$

For Dirac states (19), by using inversed Eq.(21), we obtain for $\langle T(\Psi, \overline{\Psi}) \rangle$ (26):

$$\alpha = \mu, \quad \beta = 1, \quad \lambda = \sqrt{M^2 + \mu^2}. \tag{27}$$

Since $\mu$ in this problem is an external parameter specified by spontaneous violation $\langle \Phi \rangle$, only $M$ can be defined by solution matching in the parametrization (26)-(27) (see Eq.(34)).

Even in a simplified form, the system of equations remains too complicated for qualitative analysis. In the seesaw scenario, however, we have $\alpha = \mu \ll \lambda$. It is possible to solve this problem while neglecting $\alpha$ in the numerators of the $G(p)$ propagators. In principle (if $T$ is known) we can then calculate corrections $(\alpha/\lambda)^2$. Corrections $\alpha/\lambda$ are absent. The analysis becomes then less complicated since chirality can now be altered only by $H$ interactions. An even number of vertices on each of the fermion lines does not affect chirality: $RR \to RR$, $LL \to LL$. An odd number of vertices changes chirality: $RR \rightleftharpoons LL$. A different parity of the number of vertices on two fermion lines is impossible in $T$ because $f$ preserves chirality of incoming-outgoing particles.

Let us denote the bare Majorana mass in (1) as $M_0$. The mass $M_0$ (the coefficient in the $\Psi_R$ terms, $M_0 \equiv M_R$) of course does not violate the $SU(2)$ invariance and can be present immediately in the fundamental Lagrangian. Coupled equations for $f_R$ and $f_L$ then have the form ($T$ stands for Tr$T$):

$$f_R(p) = \frac{1}{2}M_0 + \int \frac{d^4k}{(2\pi)^4 i} \frac{(+k^2)}{(\lambda^2 - k^2)^2} \Big[T_{RR}(p-k)f_R(k) + T_{RL}(p-k)f_L(k)\Big],$$
$$f_L(p) = \int \frac{d^4k}{(2\pi)^4 i} \frac{(+k^2)}{(\lambda^2 - k^2)^2} \Big[T_{LR}(p-k)f_R(k) + T_{LL}(p-k)f_L(k)\Big]. \tag{28}$$

In Eq.(28) it is possible to change over to the Euclid metrics ($-k^2 \to k^2$). Then the factors $f$, $T$ are real.

The functions $T$ do not depend on chirality, but $T_+ = T_{RR} = T_{LL}$ does not equal $T_- = T_{RL} = T_{LR}$. There is an even number of $H$ vertices on each fermion line in $T_+$ and odd one in $T_-$. Let us denote

$$f_- = f_R - f_L. \tag{29}$$

For $f_-$, we have the following equation (in Euclid metrics: $k^2 \to -k^2$, $d^4k \to -id^4k$):

$$f_-(p) = \frac{1}{2}M_0 + \int \frac{d^4k}{(2\pi)^4} \frac{k^2}{(\lambda^2 + k^2)^2} \Big[T_+(p-k) - T_-(p-k)\Big] f_-(k). \tag{30}$$

The quantity $f_-$ and its equation are selected because the equation kernel in this case represents an alternating series where terms with even and odd numbers of $H$ vertices on each of the fermion lines have opposite signs, Fig. 1.

Eq.(30) has no solution if $T$ is limited to only few diagrams with $H$-lines. At large momenta, these contributions at $p^2 \to \infty$ are asymptotically equivalent to

$$T_N(p) \sim \frac{1}{p^2} \ln^n \frac{p^2}{\lambda^2}. \tag{31}$$

The number $n < (N-1)$; $N$ is the number of $H$-lines: $T_N \sim (\mu^2/\eta^2)^N$. In this perturbative form, the equation is reduced to a differential one with boundary conditions (analogous to [7,8]). Its solution decreases: $f_-(p) \to 0$ at $p^2 \to \infty$. The integral in Eq.(30) of such a solution converges and also decreases at $p^2 \to \infty$, which contradicts the equation itself, where $f_-(p) \to \frac{1}{2}M_0$. At small Yukawa constants, the Dirac neutrino (at $M_R \neq 0$) is not possible.

The solution (30) could exist if the infinite series of $T_+ - T_-$ terms decreased at $k^2 \to \infty$ faster than $1/k^2$. With the asymptotic (31) for any single terms, this scenario is possible only if the $T_+ - T_-$ series is alternating in sign at large $k^2$. Alternating series asymptotics must be less than (31).

It is difficult to estimate the sum of diagram contributions with certain even and odd numbers of $H$-lines in $T_+$ and $T_-$. However, there is no reason to expect that even contributions have regularly different signs than odd contributions. Therefore, to generate an alternating series, the function $f_-$ is a more preferable value. Then, the existence of the solution $f(k^2 \to \infty) \to \frac{1}{2}M_0$ is connected and dependent on the convergence of the integral:

$$\left| \int \frac{d^4k}{(2\pi)^4} \frac{k^2\left(T_+(p-k) - T_-(p-k)\right)}{(\lambda^2 + k^2)^2} \right| < \infty. \tag{32}$$

The value of $f_-$ defines the Majorana mass just in the effective Lagrangian (16) for Dirac heavy neutrinos. Indeed, we have

$$\begin{aligned}
\bar{\Psi}_R^C \Psi_R f_R + \bar{\Psi}_L^{(C)} \Psi_L f_L &= \frac{1}{2}\left(\bar{\Psi}_R^C \Psi_R - \bar{\Psi}_L^C \Psi_L\right)(f_R - f_L) + \\
&+ \frac{1}{2}\left(\bar{\Psi}_R^C \Psi_R + \bar{\Psi}_L^C \Psi_L\right)(f_R + f_L) = \frac{1}{2}(\bar{\Psi}^C \gamma_5 \Psi)f_- + \frac{1}{2}(\bar{\Psi}^C \Psi)f_+.
\end{aligned} \tag{33}$$

$M$ is the solution of the equation:

$$f_-(p^2)\Big|_{p^2 = -(M^2+\mu^2)} = M, \tag{34}$$

since the Lagrangian (16) describes particles with the mass $\lambda = (M^2 + \mu^2)^{1/2}$. It is obvious from Eq.(30) that $f$ is proportional to $M_0$. The quantity $f(p^2 = -(M^2 + \mu^2))$ is real as the mass singularities of the $p^2$ for diagrams composing $f$ are higher than the quantity $p^2 > ((M^2 + \mu^2)^{\frac{1}{2}} + m_H)^2$ if $p^2$ is in non-Euclidian metrics.

With the asymptotic (31) for single terms, convergence of the integral (32) is possible only if this series is alternating. Such a scenario appears to be more achievable in the $f_-$ case, which is alternating by definition. This corresponds to the Lagrangian for Dirac neutrinos, in which case there is no solution for $f_+$ (otherwise this would have been Majorana neutrinos).

The auxiliary function $f_-(p)$ per se does not describe any process, however, if its existence is possible, then the effective Lagrangian will contain the combination $\bar{\Psi}^C \gamma_5 \Psi$ with the coefficient (34), related with the presence of only $M_R$.

The solution (34) for $f_-$ means lepton number non-conservation in $H$ boson interactions with heavy neutrinos, as seen from Eq.(23). A similar phenomenon is observed in weak interactions of these particles (see Eq.(24)).

## 4. Coupling with Light Neutrinos

In [3,4] heavy neutrinos $\Psi$ are the mirror reflected part of the common system with light fermions $\psi$. This means that their properties (apart from the mass) are only different in that $\Psi_R$ is an $SU(2)$-doublet and forms the right-handed current in weak interactions of heavy mirror particles, and $\Psi_L$ is a weak singlet.

Heavy fermions are directly related with the light part ($\psi$) of the system by terms governing transitions between the two:

$$A\bar{\psi}_L \Psi_R + B\bar{\psi}_R \Psi_L + c.c. \qquad (35)$$

Apparently, this is an $SU(2)$ invariant. Such a construction results from the system symmetrical for transpositions $\psi \rightleftharpoons \Psi$, where the transition coefficients $A$ and $B$ are the masses of the mirror-symmetrical Dirac isospinor and isoscalar ($\psi_{R,L} + \Psi_{L,R}$) included in the Lagrangian of the system prior to mirror symmetry ($\psi \rightleftharpoons \Psi$) violation. Mirror symmetry is violated by analogs of Higgs scalars.

Light neutrino masses produced by means of transitions (35) through heavy mirror states become equal to[1]:

$$m_\nu \simeq \frac{AB}{\mu}\left(\frac{\mu}{M}\right)^2. \qquad (36)$$

where $\mu$ and $M$ continue to be, respectively, Dirac and Majorana parts of the mass $\Psi$ as in (18). Eq.(36) provides direct evidence of the possibility of an exceptionally small neutrino mass $A \ll \mu \ll M$, and this evidence is more substantial than that provided by the conventional seesaw mechanism.

Light particle interactions with the Higgs scalar $\Phi$ and with the boson $H$ (both are responsible for the appearance of the parameter $\mu$) occur in a complicated manner, only via transitions into heavy $\Psi$ (see Appendix I in [9]). Transitions (35) result in the appearance of heavy $\Psi$ representatives in the physical functions of light physical neutrinos:

$$\psi_\nu \simeq \psi + \frac{A}{M}\Psi. \qquad (37)$$

---

[1] For Dirac light particles $A = B$, see [3], corrigendum in Yad.Fiz. or arXiv.

Then light neutrino interactions with $H$ appear to be proportional to the mass $m_\nu$, i.e., similar to SM, we have from (36) for the Yukawa constant

$$f \sim \frac{\mu}{\eta} \frac{AB}{M^2} \simeq \frac{m_\nu}{\eta}. \tag{38}$$

As a result, the mirror model for light neutrinos preserves the properties of the SM:

- Perturbative cancellation of Goldstone contributions with the $q^2$ pole of the gauge-invariant $W$-boson propagator;
- No fast growth of cross-sections of processes involving longitudinally polarized $W$-bosons;
- No observable lepton number violation in processes involving only light neutrinos.

Papers [3,4], as well as [9], describe these phenomena for a general case of three fermion families.

## 5. Conclusion

The primary reason for selecting the mirror mechanism with Dirac neutrinos is that it allows natural reproduction of observed qualitative WMM properties both for quarks (the CKM matrix) and leptons (the PMNS matrix).

Lepton number non-conservation with heavy Dirac neutrinos might be useful in building leptogenesis models, which have been much discussed lately (e.g., [10]), if observed neutrinos are found to be Dirac.

**Figures**

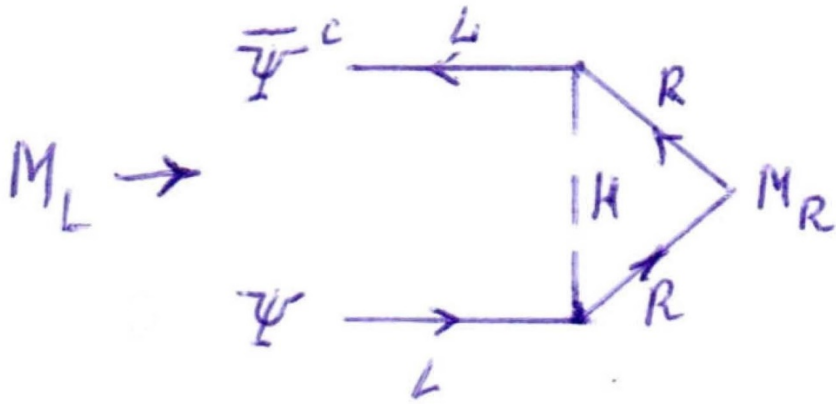

Fig.1: $M_L$ mass formation through the mass $M_R$

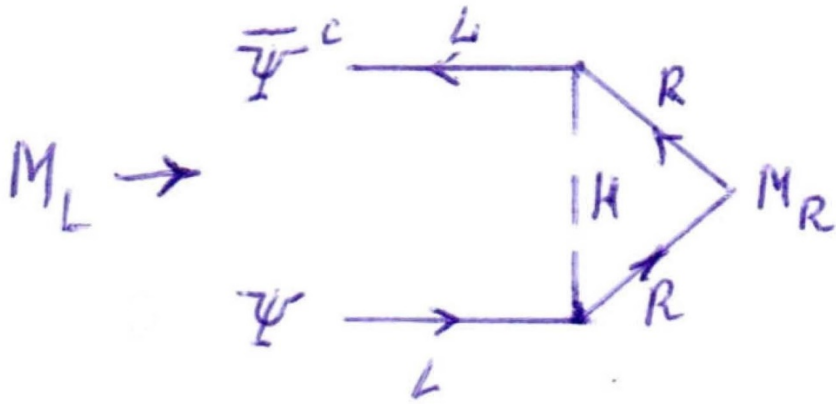

Fig.2: Equation for the coefficient $f(p)$, Eq. (25)